\documentstyle[twocolumn,seceq,letter,epsf]{jpsj}
\def\runtitle{Chiral Optical Absorption by a Single Vortex in $p_x$$\pm$i$p_y$-Wave Superconductor}
\def\runauthor{Masashige {\sc Matsumoto} and Manfred {\sc Sigrist}}

\title
{
Chiral Optical Absorption by a Vortex in $p_x$$\pm$i$p_y$-Wave Superconductor
}

\author
{ 
Masashige {\sc Matsumoto} and Manfred {\sc Sigrist}$^1$
}

\inst
{
Department of Physics, Faculty of Science, Shizuoka University, 836 Oya,
Shizuoka 422-8529, Japan \\
$^1$Yukawa Institute for Theoretical Physics, Kyoto University, Kyoto 606-8502, Japan
}

\recdate
{}
\abst
{
The chiral optical absorption by a single vortex in a
$p_x$$\pm$i$p_y$-wave superconductor 
is studied theoretically.
The $p_x$$\pm$i$p_y$-wave state was recently suggested as the symmetry
of the order parameter  
of Sr$_2$RuO$_4$ superconductor.
Due to the violation of time reversal symmetry,
there are two types of vortices whose winding orientation is the same
or opposite to the angular momentum of the Cooper pair. 
In a real material domains with $p_x$$\pm$i$p_y$-wave states are expected.
However, optical absorption of circular polarized light
depends only on the winding of the vortex
and has a low energy absorption peak which results in dichroism.
Dichroism occurs if superconductivity is realized on a single Fermi
surface sheet. However, in the case of several Fermi surface sheets 
dichroism may disappear, if the both types of carriers are present,
electron-like and hole-like. 
Therefore chiral optical absorption is a possible experiment
to detect the orbital dependent superconductivity
which was suggested as the superconducting state of Sr$_2$RuO$_4$.
}

\kword
{
Sr$_2$RuO$_4$, orbital dependent superconductivity, $p_x$$\pm{\rm i}p_y$-wave superconductor,
vortex, chiral optical absorption, dichroism, domain
}
\begin{document}
\sloppy
\maketitle
%%%%%%%%%%%%%%%%%%%%%%%%%%%%%%%%%%%%%%%%%%%%%%%%%%%%%%%%%%%%%%%%%%%%%%%%%%%%%%%%
\renewcommand{\theequation}{\arabic{equation}}
\newcommand{\no}{\noindent}
\newcommand{\beq}{\begin{equation}}
\newcommand{\beqn}{\begin{eqnarray}}
\newcommand{\eeq}{\end{equation}}
\newcommand{\eeqn}{\end{eqnarray}}
\newcommand{\ri}{{\rm i}}
\newcommand{\fx}{{\rm F}_x}
\newcommand{\vfx}{v_{{\rm F}x}}
\newcommand{\vfy}{v_{{\rm F}y}}
\newcommand{\vfxa}{v_{{\rm F}x1}}
\newcommand{\vfxb}{v_{{\rm F}x2}}
\newcommand{\delx}{{\partial \over \partial x}}
\newcommand{\dely}{{\partial \over \partial y}}

\newcommand{\br}{\mbox{\boldmath$r$}}
\newcommand{\bA}{\mbox{\boldmath$A$}}
\newcommand{\be}{\mbox{\boldmath$e$}}
\newcommand{\rd}{{\rm d}}
\newcommand{\bk}{\mbox{\boldmath$k$}}
\newcommand{\kh}{{\hat k}}
\newcommand{\tk}{{\theta_k}}
\newcommand{\kf}{k_{\rm F}}
\newcommand{\kfx}{{k_{{\rm F}x}}}
\newcommand{\kfxa}{{k_{{\rm F}x1}}}
\newcommand{\kfxb}{{k_{{\rm F}x2}}}
\newcommand{\bkf}{{\mbox{\boldmath$k$}}_{\rm F}}
\newcommand{\om}{{\omega_{m}}}
\newcommand{\Ps}{{\hat \Psi}}
\newcommand{\Ph}{{\hat \Phi}}
\newcommand{\Pha}{\Ph_\alpha}
\newcommand{\Phb}{\Ph_\beta}
\newcommand{\Del}{{\hat \Delta}}
\newcommand{\Dela}{\Delta_\alpha}
\newcommand{\Delb}{\Delta_\beta}
\newcommand{\Delah}{{{\hat \Delta}_\alpha}}
\newcommand{\Delbh}{{{\hat \Delta}_\beta}}
\newcommand{\bDelta}{{\bar \Delta}}
\newcommand{\tauc}{{\hat \tau_3}}
\newcommand{\taub}{{\hat \tau_2}}
\newcommand{\G}{{\hat G}}
\newcommand{\g}{{\hat g}}
\newcommand{\ab}{{\alpha \beta}}
\newcommand{\U}{{\hat U}}
\newcommand{\TU}{{\tilde U}}
\newcommand{\Ua}{{{\hat U}_\alpha}}
\newcommand{\Ub}{{{\hat U}_\beta}}
\newcommand{\A}{{\hat A}}
\newcommand{\B}{{\hat B}}
\newcommand{\Lam}{{\hat \Lambda}}
\newcommand{\re}{{\rm e}}
\newcommand{\bd}{\mbox{\boldmath{$d$}}}
\newcommand{\bR}{\mbox{\boldmath{$R$}}}
\newcommand{\bX}{\mbox{\boldmath{$X$}}}
\newcommand{\hDelta}{\hat \Delta}
\newcommand{\hpsi}{\hat \psi}
\newcommand{\hPhi}{\hat \Phi}
\newcommand{\hG}{\hat G}
\newcommand{\hF}{\hat F}
\newcommand{\bq}{\mbox{\boldmath{$q$}}}
\newcommand{\hsigma}{\hat \sigma}
\newcommand{\htau}{\hat \tau}
\newcommand{\bsigma}{\hat {\mbox{\boldmath{$\sigma$}}}}
\newcommand{\bn}{\boldmath{\mbox{$n$}}}
\newcommand{\bp}{\mbox{\boldmath{$p$}}}
\newcommand{\Q}{\mbox{\boldmath{$Q$}}}
\newcommand{\bx}{\mbox{\boldmath{$x$}}}
\newcommand{\hkf}{{\hat k}_{\rm F}}
\newcommand{\vf}{v_{\rm F}}
\newcommand{\psiu}{\psi_\uparrow}
\newcommand{\psid}{\psi_\downarrow}
\newcommand{\ep}{\varepsilon}
\renewcommand{\dag}{\dagger}
\newcommand{\taua}{\hat {\tau_1}}
\newcommand{\sigb}{\hat {\sigma_2}}
\newcommand{\del}{\partial}
\newcommand{\ome}{\omega_{\rm m}}
\newcommand{\bfkp}{\mbox{\boldmath $k_+$}}
\newcommand{\bfkn}{\mbox{\bolkmath $k_-$}}
\newcommand{\bskf}{\mbox{\footnotesize \boldmath $k_{\rm F}$}}
\newcommand{\bsq}{\mbox{\footnotesize \boldmath $q$}}
\newcommand{\bsr}{\mbox{\footnotesize \boldmath $r$}}
\newcommand{\bfsq}{\mbox{\footnotesize \boldmath $q$}}
\newcommand{\bfskp}{\bfsk_+}
\newcommand{\bfskn}{\bfsk_-}
\newcommand{\bfsqp}{\bfsq_+}
\newcommand{\bfsqn}{\bfsq_-}
\newcommand{\rhok}{\rho_\bfsk}
\newcommand{\rhokp}{\rho_\bfsk}
\newcommand{\xperp}{x_\perp}
\newcommand{\ypara}{y_\parallel}
\newcommand{\ky}{k_y}
\newcommand{\m}{\hat \mu}
\newcommand{\n}{\hat \nu}
\newcommand{\phai}{\hat \varphi}
\newcommand{\pha}{\hat \phi}
\newcommand{\qy}{q_y}
\newcommand{\sig}{\hat \Sigma}
\newcommand{\tsigb}{\Sigma_2}
\newcommand{\tsigc}{\Sigma_3}
\newcommand{\fbar}{\overline{f}}
\newcommand{\dbar}{\overline{D}}
\newcommand{\fig}[1]
{
\vspace{24pt}
\begin{center}
\fbox{\rule{0cm}{#1}\hspace{7cm}}
%\fbox{\rule{0cm}{7cm}\hspace{7cm}}
\end{center}
}
%%%%%%%%%%%%%%%%%%%%%%%%%%%%%%%%%%%%%%%%%%%%%%%%%%%%%%%%%%%%%%%%%%%%%%%%%%%%%%%%
The study of unconventional superconductivity has become one of the most
attractive problems in recent condensed matter research, 
since various examples of this class have been discovered among 
strongly correlated electron systems. They include some
heavy fermion compounds and the high-temperature superconductors
and the more recently discovered Sr$_2$RuO$_4$.
\cite{Maeno}

It has been suggested early that Sr$_2$RuO$_4$ might be a spin triplet
$p$-wave superconductor, in particular, due to its relation to various ferromagnetic compounds.
\cite{Gibb,Rice}
Meanwhile a considerable bulk of experimental evidence has been collected
supporting this theoretical proposal.
For example the absence of a Hebel-Slichter peak in NQR
\cite{Ishida}
and the sensitivity of $T_{\rm C}$ on non-magnetic impurities
\cite{Mackenzie2}
clearly point towards unconventional pairing. 
Moreover, the indication of broken time reversal symmetry in the
superconducting phase, observed in $\mu$SR measurements, gives a 
strong argument for the $p$-wave symmetry.
\cite{Luke}
The most decisive clue for spin-triplet $p$-wave pairing comes from the Knight shift
experiment which shows that the spin susceptibility is not affected by 
the superconducting state.
\cite{Ishida2}
The most likely pairing state with broken of the time reversal symmetry
in a tetragonal crystal field is given by $\bd(\bk)$=$(k_x$$\pm$$\ri k_y){\hat z}$,
the $p_x$$\pm$i$p_y $-wave state.

From previous investigations, it is however know that dichroism for the bulk states
like the $p_x$$\pm$i$p_y$-wave state are probably small.
\cite{Li,Sauls}
Hence we will concentrate here on the inhomogeneous regions of the superconductor 
the vortex and the surface.
For the $s$-wave case, dichroism in the chiral optical absorption by a
single vortex was studied previously.
\cite{Zhu,Janko}
In this paper we extend this type of study to the $p_x$$\pm$i$p_y$-wave pairing state. 

%%%%%%%%%%%%%%%%%%%%%%%%%%%%%%%%%%%%%%%%%%%%%%%%%%%%%%%%%%%%%%%%%%%%%%%%%%%%%%%%
For simplicity, we assume that the superconductor is basically two
dimensional and has a cylindrical Fermi surface.
We then apply the method developed for the study of vortices in an $s$-wave superconductor.
\cite{Gygi,Hayashi}
Let us start with the following Bogoliubov de-Gennes equation,
\cite{Morita}
\beqn
&&h_0 u_n(\br)-{\ri \over \kf}\Bigl\{\Delta(\br)\Box^+
+{1 \over 2}\bigl[\Box^+\Delta(\br)\bigr]\Bigr\}v_n(\br)=E_n u_n(\br), \cr
&&-h_0v_n(\br)-{\ri \over \kf}\Bigl\{\Delta^*(\br)\Box^-
+{1 \over 2}\bigl[\Box^-\Delta^*(\br)\bigr]\Bigr\}u_n(\br)=E_n v_n(\br). \nonumber \\
\label{eqn:BdG}
\eeqn
where $h_0$=$-$$\nabla^2/2m$$-$$E_{\rm F}$,
$\Box^\pm$=$\partial/\partial x$$\pm$i$\partial/\partial y$,
$\Delta(\br)$ is the order parameter,
$\kf$ is the Fermi wave number,
$E_n$ is the $n$-th excitation energy (positive value) and $E_{\rm F}$
is the Fermi energy ($\hbar$=1).
The wave functions $u_n$ and $v_n$
describe the quasiparticle-hole spinor state of energy $ E_n $.
Here the vector potential is neglected and this is valid for strong type-II superconductors.
Though Sr$_2$RuO$_4$ is not of this type,
we use eq. (\ref{eqn:BdG}) for simplicity.
There is a relation between the solutions of $E_n$ and $-$$E_n$, 
\beq
\bigl\{
  u_{-E_n}(\br),~v_{-E_n}(\br)
\bigr\}
\leftrightarrow
\bigl\{
  v_{E_n}^*(\br),~u_{E_n}^*(\br)
\bigr\} .
\label{eqn:symmetry}
\eeq
The vortex center shall be located at the origin of the coordinate.
Then the order parameter can be written as
$\Delta(\br)$=$\bDelta(\br){\rm e}^{\ri\alpha\theta}$,
where $\alpha$=$\pm 1$ represents the direction of the vortex (positive or
negative winding) and $\theta$ is the angle of $\br$ measured relative 
to $x$-axis. Furthermore,
$\bDelta(\br)$ is the magnitude of the order parameter. 
In conventional superconductors the winding orientation of the vortex
does not affect its properties apart from the direction of the
magnetic flux. This is different in the case 
of the $p_x$+i$p_y$-wave superconductor, since it breaks 
time reversal symmetry and introduces its own winding. In the
following we will restrict to one of the two degenerate states, the $p_x$+i$p_y$-wave states
and neglect $p_x$$-$i$p_y$-wave part which is induced near the vortex core for simplicity.
\cite{Agterberg}
All results can be generalized
through the time reversal operation.

\begin{figure}
\begin{center}
\begin{minipage}[h]{8cm}
\epsfxsize=8cm
\epsfbox{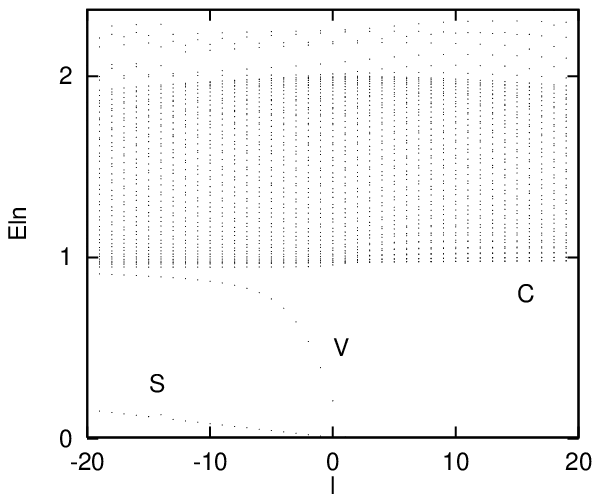}
\end{minipage}
\begin{minipage}[h]{5cm}
\epsfxsize=5cm
\epsfbox{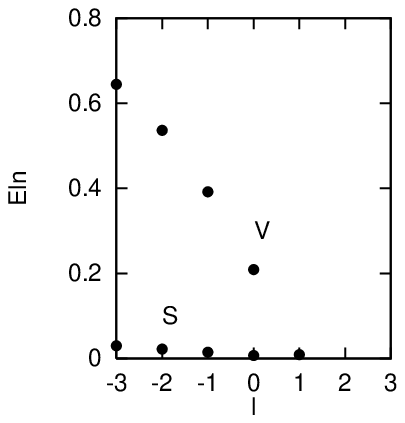}
\end{minipage}
\end{center}
\caption{
Excitation energies normalized by $\Delta_0$ for the negative vortex.
S, V and C represent the surface bound state, vortex bound state and continuum state,
respectively.
Set of parameters are chosen as
$R$=10$\pi\xi$, $\kf\xi$=4, $\omega_{\rm C}$=2$\Delta_0$.
%Here $\kf$ is the Fermi wave number.
$R$=10$\pi\xi$ corresponds to the level discreetness in the normal state
as $\delta\epsilon$$\simeq$0.1$\Delta_0$.
S and V states close to zero angular momentum are also shown.
}
\label{fig:E1}
\end{figure}

First we address the vortex with positive winding.
We solve eq. (\ref{eqn:BdG}) in a finite size system of circular shape with radius is $R$.
It is convenient to expand $u_n$ and $v_n$ as
\beqn
\left(
  \matrix{
    u_n(\br) \cr
    v_n(\br) \cr
  }
\right)
&=&
{1 \over \sqrt{2\pi}}{\displaystyle \sum_j}
\left(
  \matrix{
    u_{lnj}\re^{\ri l\theta}\varphi_{lj}(r) \cr
    v_{l-2nj}\re^{\ri (l-2)\theta}\varphi_{l-2j}(r) \cr
  }
\right), \cr
\varphi_{lj}(r)&=&{\sqrt{2} \over R\vert J_{l+1}(Z_{lj})\vert}J_l({Z_{lj}r \over R}),
\label{eqn:expand1}
\eeqn
where $J_l$ is the $l$-th Bessel function, $Z_{lj}$ is the $j$-th zero of $J_l$. 
Substituting eq. (\ref{eqn:expand1}) into eq. (\ref{eqn:BdG}),
we obtain,
\beqn
&&\epsilon_{lj}u_{lnj}-\ri{\displaystyle \sum_{j'}}\bDelta^{+-}_{ljl-2j'}v_{l-2j'}
= E_{ln} u_{lnj}, \cr
&&-\epsilon_{l-2j}v_{l-2nj}-\ri{\displaystyle \sum_{j'}}\bDelta^{++}_{l-2jlj'}u_{lj'}
= E_{ln} v_{l-2nj}, \cr
&&\bDelta^{\alpha\pm}_{ljl'j'}
=
{1 \over \kf}\int_0^R \rd r
r\varphi_{lj}(r)\Bigl\{
  \bDelta(r)({\del \over \del r}\pm {l' \over r}) \cr
&&~~~~~~~~~~~~~~~~~~~~+{1 \over 2}\bigl[({\del \over \del r}-{\alpha \over r})\bDelta(r)\bigr]
\Bigr\}\varphi_{l'j'}(r),
\label{eqn:BdG1}
\eeqn
where $\epsilon_{lj}$=$(Z_{lj}/R)^2/2m$$-$$E_{\rm F}$, and 
$E_{ln}$ is the $n$-th excitation energy with $l$
which is the angular momentum index referred to that of the spinor component $u_l$.
It is sufficient for our purpose to approximate the form of the order
parameter by $\bDelta(r)$=$\Delta_0{\rm tanh}(r/\xi)$ in order to avoid the 
complications of a full self-consistence calculation
($\xi$=$\vf/\Delta_0$: coherence length; $\vf$: Fermi velocity).
It is important to notice that in eq. (\ref{eqn:BdG1}) 
$u_l$ couples with $v_{l-2}$ only.
Without the vortex, the coupling is between $u_l$  and $v_{l-1}$
due to the internal angular momentum of the  $p_x$+i$p_y$-wave state.
It is the phase winding of the order parameter around the vortex which makes the difference. 
Introducing a cut off energy $\omega_{\rm C}$,
we can diagonalize eq. (\ref{eqn:BdG1}) and obtain
the excitation energies as shown in Fig. \ref{fig:E1}.
The set of parameters used here corresponds to the quantum limit case since $\kf\xi$=4,
however,
the essential properties hold if the system is close to the classical limit (large $\kf\xi$).
There are two kinds of bound states, 
one localized around the vortex (vortex bound state)
and the other formed near the surface at $r$=$R$ (surface bound state).
The latter only occurs for anisotropic pairing.
\cite{Matsumoto}
The vortex and surface bound states exist only for $l$$\le$1.
Due to eq. (\ref{eqn:symmetry})
the bound state with angular momentum index $l$=1 contains
both vortex and surface bound states features.

\begin{figure}
%\epsfile{file=stsp-vortex.eps}
%\epsfile{file=stsp-surface.eps}
\begin{center}
\begin{minipage}[t]{9cm}
\epsfxsize=9cm
\epsfbox{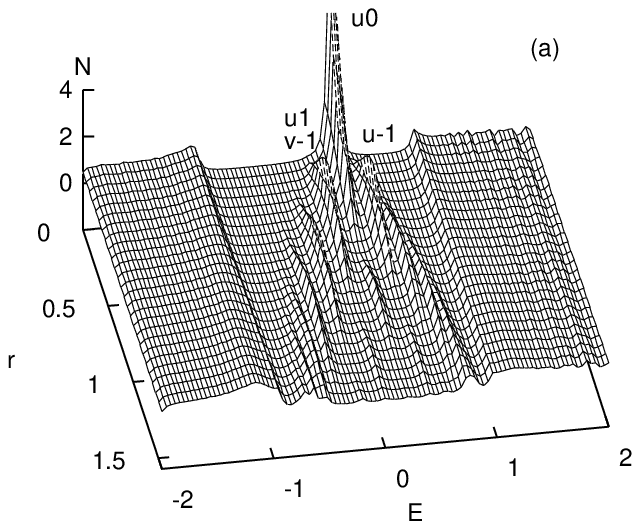}
\end{minipage}
\begin{minipage}[t]{9cm}
\epsfxsize=9cm
\epsfbox{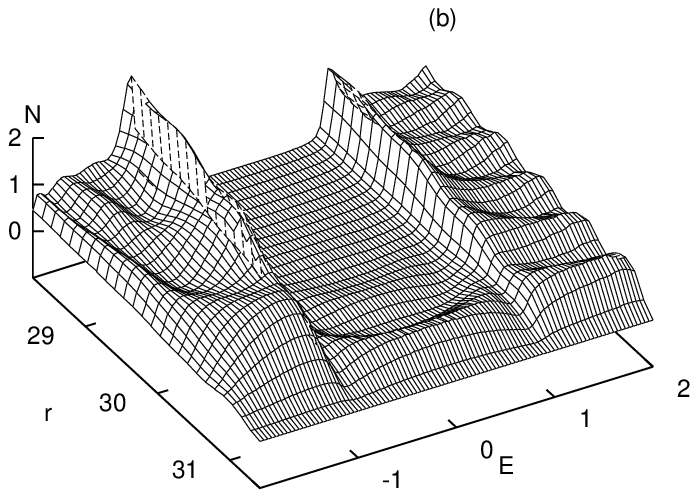}
\end{minipage}
\end{center}
\caption{
Local density of states for the positive vortex in an arbitrary unit.
A small imaginary part of 0.03$\Delta_0$ is added to $E$ for convenience.
(a)~Near the vortex core.
(b)~Near the surface.
In each figure $r$ and $E$ are normalized by $\xi$ and $\Delta_0$, respectively.
}
\label{fig:dos1}
\end{figure}

One tool to investigate the quasiparticle states in the superconductor 
is scanning tunneling microscopy. This technique probes basically 
the local density of states,
\beq
N(\br,E)={\displaystyle \sum_n}
\Bigl[
  \vert u_n(\br)\vert^2\delta(E-E_n)
 +\vert v_n(\br)\vert^2\delta(E+E_n)
\Bigr]
\eeq
as shown in Fig.\ref{fig:dos1}.
Near the vortex the bounds states appear as distinct subgap peaks in the density of states.
%We can identify what bound states contributes to the peaks.
We label the bound state peaks with corresponding $u_l$ and $v_l$.
This feature is very similar to the behavior found for $s$-wave superconductors. 
Different to the $s$-wave case, however, we find here that
bound states yield a local density of states similar to the normal
metal close to the surface ($r$=$R$) as shown in Fig. \ref{fig:dos1}(b).
\cite{Matsumoto}
We observe here also the feature of Friedel
oscillations which have a wave vector $2k_F$. 

Next we turn to the opposite vortex (negative winding).
The Bogoliubov de-Gennes equation takes the form,
\beqn
&&\epsilon_{lj}u_{lnj}-\ri{\displaystyle \sum_{j'}}\bDelta^{--}_{ljlj'}v_{lj'}
= E_{ln} u_{lnj}, \cr
&&-\epsilon_{lj}v_{lnj}-\ri{\displaystyle \sum_{j'}}\bDelta^{-+}_{ljlj'}u_{lj'}
= E_{ln} v_{lnj},
\label{eqn:BdG2}
\eeqn
where $u_n(\br)$ and $v_n(\br)$ have been expanded as
\beqn
\left(
  \matrix{
    u_n(\br) \cr
    v_n(\br) \cr
  }
\right)
&=&
{1 \over \sqrt{2\pi}} \re^{\ri l\theta}{\displaystyle \sum_j}
\left(
  \matrix{
    u_{lnj} \cr
    v_{lnj} \cr
  }
\right)
\varphi_{lj}(r).
\label{eqn:expand2}
\eeqn
In this case $u_l$ couples with $v_l$. 
Naturally the surface bound states appear for $l$$\le$0 analogous to the previous case.
However, the vortex bound states are now restricted 
the angular momenta $l$$\ge$0, reversed to the vortex with negative winding.
The bound state with index $l$=0 contains both a vortex and a surface bound state features.

\begin{figure}
%\epsfile{file=stsn-vortex.eps}
%\epsfile{file=stsn-surface.eps}
\begin{center}
\begin{minipage}[t]{9cm}
\epsfxsize=9cm
\epsfbox{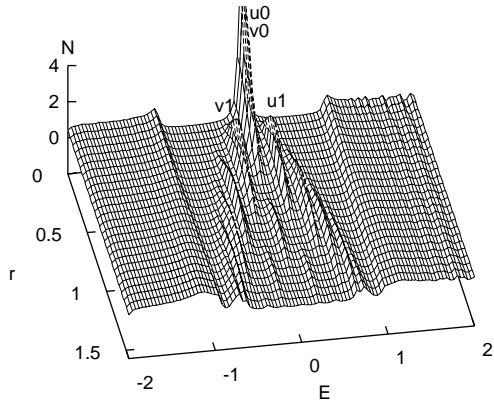}
\end{minipage}
\end{center}
\caption{
Local density of states near the vortex core for the negative vortex in an arbitrary unit.
A small imaginary part of 0.03$\Delta_0$ is added to $E$ for convenience.
$r$ and $E$ are normalized by $\xi$ and $\Delta_0$, respectively.
}
\label{fig:dos2}
\end{figure}

\begin{figure}
\begin{center}
\begin{minipage}[t]{7cm}
\epsfxsize=7cm
\epsfbox{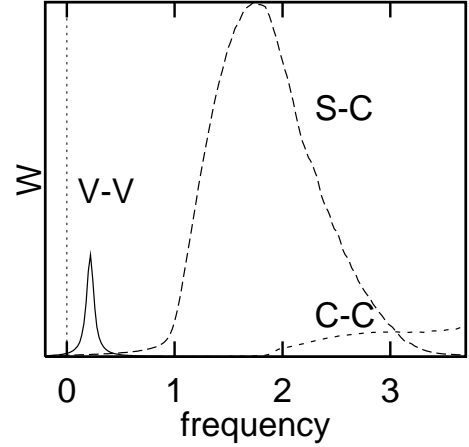}
\end{minipage}
\end{center}
\caption{
Optical absorption rate of the left circular polarized light by a positive vortex
for the $p_x$+i$p_y$-wave order parameter in an arbitrary unit.
The frequency is normalized by $\Delta_0$.
A small imaginary part of $0.04\Delta_0$ is added to $\omega$ for convenience.
V-V, S-C and C-C represent the absorption caused by the excitations of
vortex-vortex, surface-continuum and continuum-continuum states, respectively.
The total absorption can be obtained by adding these three with the same weight.
}
\label{fig:abs}
\end{figure}

We now consider the difference in the physical properties between the two vortices.
In Fig. \ref{fig:dos2} we show the local density of states
only near the vortex core for the negative winding vortex,
since the winding of vortex does not affect it near the surface.
The peak near the vortex core comes from zero angular momentum,
since only $J_0$ has a finite value at $r$=0.
For the negative vortex $u_0$ couples with $v_0$
and they have the lowest vortex bound state
which makes a peak at almost zero energy near the core.
On the other hand,
$u_2$ couples with $v_0$ for the positive vortex case
and it forms a peak near the core at the second lowest vortex bound state energy
(see Fig. \ref{fig:dos1}).
Increasing $\kf\xi$ (classical limit) the difference becomes small
due to the reduced spacing of the bound state energy level.
Because the vortex bound states are associated with specific angular momenta,
we expect, however, to see a distinction in the chiral optical absorption by the vortices.
The vector potential of the circular polarized electromagnetic
radiation is described by 
$\bA^\pm(\br)$=$A_q(\be_x$$\pm$i$\be_y){\rm exp}[\ri(\bq\cdot\br$$-$$\omega t)]$,
where $\be_x$ and $\be_y$ are the unit vector of the $x$ and $y$ axes,
respectively. 
The superscripts $^\pm$ denotes the right and left circular
polarization, respectively. 
We assume that $\bq$ is a small wave vector in the $z$-direction.
The absorption rate $W^\pm$ at $T$=0 
can be expressed as
\beqn
W^\pm&=&2\pi({e \over m})^2{\displaystyle \sum_{nn'}}
\vert M_{nn'}^\pm\vert^2\delta(E_n+E_{n'}-\omega), \cr
M_{nn'}^\pm&=&A_q\int\rd\br
\bigl[
 u_n^*(\br)\Box^\pm v_{n'}^*(\br)-u_{n'}^*(\br)\Box^\pm v_n^*(\br)
\bigr]. \nonumber \\
\label{eqn:abs}
\eeqn
where $e$ is the electron charge.
Since $\Box^\pm$ has the form of
$\re^{\pm\ri\theta}({\del \over \del r}$$\pm$i${1 \over r}{\del \over
\del\theta})$, 
the angular momentum of $u_n$ and $v_{n'}$ must satisfy 
the rule $l_u$+$l_v$=$\pm$1 (conservation of the angular momentum).
The absorption of radiation causes the excitation of two quasiparticles.
In Fig. \ref{fig:abs} we show the absorption rate of the left polarized light
for the vortex with positive winding.
We show only three types of absorptions,
because the weight of other processes are very small.
The main absorption edge lies at $\omega/\Delta_0$$\simeq$1,
which does not depend on the vortex winding orientation nor the polarization of the radiation.
This absorption is mainly caused by the excitations of a surface bound state and a continuum state,
so that it would also appear without the vortex.
Nevertheless, due to the fact that the surface states
are a characteristic feature of anisotropic pairing,
this low absorption edge provides evidence for non-$s$-wave pairing
as the edge to the pure continuum excitations lies at $\omega/2 \Delta_0$$\simeq$1.

\begin{figure}
\begin{center}
\input{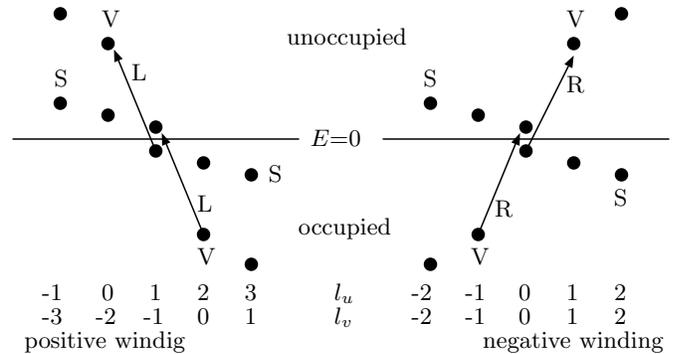}
\end{center}
\caption{
Transition from occupied to unoccupied bound states
for the right (R) and left (L) circular polarized light.
The order parameter is assumed to be $p_x$+i$p_y$-wave.
}
\label{fig:selection}
\end{figure}

There is a further characteristic point in the absorption.
There are two negligible contributions of low energy.
The vortex-surface state excitation is small
due to the long distance between the vortex core and surface.
In the surface-surface case,
the matrix element is completely zero in the absence of a vortex,
and, thus, depends also strongly on the distance between vortex and surface. 
Then the only low-energy contribution comes from the excitation of two vortex states (V-V).
With the symmetry relation eq. (\ref{eqn:symmetry})
the selection rule can be understood by Fig. \ref{fig:selection}.
For the right circular polarized light
there is no such contribution from the V-V absorption for the positive winding vortex.
The absorption features for the vortex with negative winding is almost the same,
if the polarization of the light is reversed as expressed in Fig. \ref{fig:selection}.
This means that dichroism in the absorption is strongly connected
with only the winding direction of the vortex relative to the polarization of the light.
In the real material we expect that domains of the two degenerate pairing states occur
and that both types of domains can contribute to the absorption.
Even if they exist we can observe the dichroism in the absorption
by applying the same external field to both domains in the sample,
since the selection of R or L polarized light depend only on the
winding direction of the vortex. 
This result is essentially the same as the $s$-wave case.
\cite{Zhu,Janko}

An interesting aspect occurs in connection with orbital dependent
supercondutivity. Various experiments reporting a large residual
density of states at low temperature \cite{Ishida,Nishizaki} indicate that 
the superconducting state of Sr$_2$RuO$_4$ 
is realized predominantly on the Fermi surface of the $4d_{xy}$-orbital 
\cite{Agterberg,Agterberg2}.
The other interpretation of this is given by a non-unitary state.
\cite{Machida}
If the orbital dependent superconductivity is really the case,
the relevant superconducting carriers
are electrons, 
while for the $4d_{yz}$ and $4d_{zx}$ orbitals would also contribute 
hole-like carriers, since both electron and hole Fermi surfaces  are
formed by these two orbitals.
The solutions of eq. (\ref{eqn:BdG}) for the electron and hole carriers
show the following symmetry relations:
\beq
u\leftrightarrow v,~~~\Box^\pm\leftrightarrow\Box^\mp,~~~\Delta\leftrightarrow\Delta^*.
\label{eqn:trans}
\eeq
Equation (\ref{eqn:abs}) implies that the first transformation in eq. (\ref{eqn:trans})
does not change the absorption rate. Also the second transformation
which corresponds to $p_x$+i$p_y$$\leftrightarrow$$p_x$$-$i$p_y$  does not
affect our previous result. On the other hand, the third one reverses
the winding of the vortex and the effect of polarization (R or L) such 
that the condition for the low-energy absorption peak is reversed.
\cite{Zhu}
Consequently, we expect to observe the low-energy absorption peak for
both R and L polarized light,  
if $4d_{yz}$ and $4d_{zx}$ orbitals contributes to the superconductivity.

We list the result of the appearance of the low frequency peak in Table \ref{tab:dichroism}. 
\begin{table}
\caption{
Appearance of the absorption peak at low frequency for the positive winding vortex.
R and L represent the right and left circular polarization, respectively.
}
\label{tab:dichroism}
\begin{tabular}{@{\hspace{\tabcolsep}\extracolsep{\fill}}ccc} \hline
& R  & L \\ \hline
$4d_{xy}$ orbital only & $\times$ & $\bigcirc$ \\
$4d_{yz},4d_{zx}$ orbitals included & $\bigcirc$ & $\bigcirc$ \\ \hline
\end{tabular}
\end{table}

In summary, there are features in the optical absorption which could provide evidence for
unconventional properties of superconductivity in Sr$_2$RuO$_4$. 
One is the surface bound state which yields an absorption edge at $\omega/\Delta_0$=1,
lower than the threshold of the two continuum states
excitations. 
The other is the dichroism in the absorption
caused by the excitations of two vortex bound states. The latter can
also give evidence for the orbital dependent superconductivity for Sr$_2$RuO$_4$. 
Unfortunately, the energy of the vortex bound state excitation is
rather low so 
that it would be only observable in the microwave range,
which limits the resolution drastically.
The absorption edge may, therefore, be a more promising experimental target
to probe the unconventional superconducting state of Sr$_2$RuO$_4$.

%%%%%%%%%%%%%%%%%%%%%%%%%%%%%%%%%%%%%%%%%%%%%%%%%%%%%%%%%%%%%%%%%%%%%%%%%%%%%%%%
\section*{Acknowledgments}
%%%%%%%%%%%%%%%%%%%%%%%%%%%%%%%%%%%%%%%%%%%%%%%%%%%%%%%%%%%%%%%%%%%%%%%%%%%%%%%%
We are very grateful to N. Hayashi and R. Heeb
for many helpful discussions on conceptional and technical aspects of this work.
%%%%%%%%%%%%%%%%%%%%%%%%%%%%%%%%%%%%%%%%%%%%%%%%%%%%%%%%%%%%%%%%%%%%%%%%%%%%%%%%

%%%%%%%%%%%%%%%%%%%%%%%%%%%%%%%%%%%%%%%%%%%%%%%%%%%%%%%%%%%%%%%%%%%%%%%%%%%%%%%%
\clearpage

\vspace*{5cm}

%%%%%%%%%%%%%%%%%%%%%%%%%%%%%%%%%%%%%%%%%%%%%%%%%%%%%%%%%%%%%%%%%%%%%%%%%%%%%%%%
%%%%%%%%%%%%%%%%%%%%%%%%%%%%%%%%%%%%%%%%%%%%%%%%%%%%%%%%%%%%%%%%%%%%%%%%%%%%%%%%

%%%%%%%%%%%%%%%%%%%%%%%%%%%%%%%%%%%%%%%%%%%%%%%%%%%%%%%%%%%%%%%%%%%%%%%%%%%%%%%%
%%%%%%%%%%%%%%%%%%%%%%%%%%%%%%%%%%%%%%%%%%%%%%%%%%%%%%%%%%%%%%%%%%%%%%%%%%%%%%%%

%%%%%%%%%%%%%%%%%%%%%%%%%%%%%%%%%%%%%%%%%%%%%%%%%%%%%%%%%%%%%%%%%%%%%%%%%%%%%%%%
\begin{figure}
\begin{center}
\begin{minipage}[t]{8cm}
\epsfxsize=8cm
\epsfbox{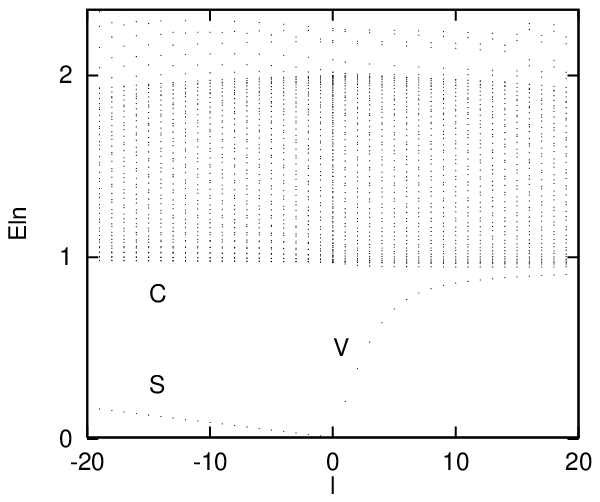}
\end{minipage}
\begin{minipage}[t]{5cm}
\epsfxsize=5cm
\epsfbox{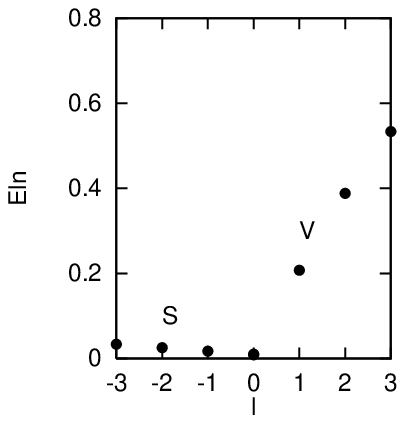}
\end{minipage}
\end{center}
\caption{
Excitation energies normalized by $\Delta_0$ for the negative vortex.
S and V states close to zero angular momentum are also shown.
}
\label{fig:E2}
\end{figure}
%%%%%%%%%%%%%%%%%%%%%%%%%%%%%%%%%%%%%%%%%%%%%%%%%%%%%%%%%%%%%%%%%%%%%%%%%%%%%%%%

%%%%%%%%%%%%%%%%%%%%%%%%%%%%%%%%%%%%%%%%%%%%%%%%%%%%%%%%%%%%%%%%%%%%%%%%%%%%%%%%
%%%%%%%%%%%%%%%%%%%%%%%%%%%%%%%%%%%%%%%%%%%%%%%%%%%%%%%%%%%%%%%%%%%%%%%%%%%%%%%%

%%%%%%%%%%%%%%%%%%%%%%%%%%%%%%%%%%%%%%%%%%%%%%%%%%%%%%%%%%%%%%%%%%%%%%%%%%%%%%%%
%%%%%%%%%%%%%%%%%%%%%%%%%%%%%%%%%%%%%%%%%%%%%%%%%%%%%%%%%%%%%%%%%%%%%%%%%%%%%%%%

%%%%%%%%%%%%%%%%%%%%%%%%%%%%%%%%%%%%%%%%%%%%%%%%%%%%%%%%%%%%%%%%%%%%%%%%%%%%%%%%
%%%%%%%%%%%%%%%%%%%%%%%%%%%%%%%%%%%%%%%%%%%%%%%%%%%%%%%%%%%%%%%%%%%%%%%%%%%%%%%%

\end{document}